\documentclass[a4paper,twoside]{article}

\usepackage{epsfig}
\usepackage{subcaption}
\usepackage{calc}
\usepackage{amssymb}
\usepackage{amstext}
\usepackage{amsmath}
\usepackage{amsthm}
\usepackage{multicol}
\usepackage{pslatex}
\usepackage{apalike}
\usepackage[bottom]{footmisc}

\usepackage{url}

\usepackage{SCITEPRESS}     

\begin{document}

\title{Data Exfiltration by Hotjar Revisited}

\author{\authorname{Libor Polčák\sup{1}\orcidAuthor{0000-0001-9177-3073} and Alexandra Slezáková}
\affiliation{\sup{1}Brno University of Technnology, Faculty of Information
Technology, Božetěchova 2, 612 66 Brno, Czech Republic}
\email{\{polcak\}@fit.vut.cz}
}


\abstract{Session replay scripts allow website owners to record the interaction of
each web site visitor and aggregate the interaction to reveal the interests and
problems of the visitors. However, previous research identified such techniques
as privacy intrusive. This position paper updates the information on data
collection by Hotjar. It revisits the previous findings to detect and describe
the changes.
The default policy to gather inputs changed; the
recording script gathers only information from explicitly allowed input
elements. Nevertheless, Hotjar does record content reflecting users' behaviour outside input HTML
elements. Even though we propose changes that would prevent the leakage of the
reflected content, we argue that such changes will most likely not appear in
practice. The paper discusses improvements in handling TLS. Not
only do web page operators interact with Hotjar through encrypted
connections, but Hotjar scripts do not work on sites not protected by TLS.
Hotjar respects the Do Not Track signal; however, users need to connect to Hotjar
even in the presence of the Do Not Track setting. Worse, malicious web operators
can trick Hotjar into recording sessions of users with the active Do Not Track setting.
Finally, we propose and motivate the extension of GDPR Art. 25 obligations to
processors.
}

\onecolumn \maketitle \normalsize \setcounter{footnote}{0} \vfill

\section{\uppercase{Introduction}}
\label{sec:introduction}

Website operators want to monitor the interaction of the visitors of the
websites, for example, to detect problems. Session recording tools allow
detailed
monitoring of user
behaviour~\cite{session_replay_comparison,session-replay,session_replay_nissenbaum}.
Consequently, session recording scripts are widespread even though users are
typically unaware of the detailed monitoring of their
behaviour~\cite{session-replay,session_replay_nissenbaum}.
Session recording scripts record users' mouse movements, clicks, and keystrokes.
The recording script sends the collected data to servers of the recording party, which
provide aggregated statistics in the form of heat maps but also entire
recordings of unique users.
Recordings often contain sensitive user data, such as medical conditions, credit
card information, and other data that can subsequently be misused~\cite{session-replay}.

\cite{session-replay} analysed six session recording companies in 2018
and revealed password, credit card, and health data leaks.
Similarly to \cite{session_replay_nissenbaum} that revisited one of the
companies, FullStory, this position paper
revisits and updates the results of the study of Acar et al. This paper focuses on Hotjar as it is one of the most popular replay service companies in the market. Acar et al. detected
Hotjar on many sites.
Even so, Acar et al. do not list any change in the data collection of
Hotjar.

In 2018, Hotjar collected (1) texts typed into forms before the user submitted the form and
(2) precise mouse movements. Both without any visual indication to the
user \cite[Section 6.1]{session-replay}. At the time of the study, Hotjar
collected the text inputs verbatim by default, except for passwords, credit card
numbers, and partially addresses.
Moreover, \cite{session-replay} detected session replay companies that transfer
the content of HTTPS websites over HTTP, not protected by any encryption.

This position paper revisits Hotjar in 2023 and reveals that Hotjar's default
behaviour changed. The content of all inputs is no longer collected by default.
Hotjar changed its policy to
collect only inputs marked with \texttt{data-hj-allow} attribute.
\cite{session-replay} argued that the opt-out from data collection is not
practical. Hotjar likely accepted the arguments.

\cite{session-replay} also warned that the displayed content on a page could
reflect user-specific content\footnote{Reflected information is any information
initially
gathered from a user that is subsequently propagated to the DOM or HTML at a
different position.}.
For example, online stores typically show the
content of the user basket and the filled billing information as a part of the
page just before checkout. Our results show that Hotjar's scripts gather the
content of all elements except inputs
by default, so the user-specific information leaks the reflected
information.

The general shift to encrypted HTTP connections motivated Hotjar to
operate through HTTPS. Moreover, Hotjar refuses to work on HTTP sites.
Even more, we show that Hotjar
respects the do not track settings signalled by the browser. However, we also tested
that malicious sites can force Hotjar to record users who opt out of
tracking.

\cite{session-replay} observed that by bringing attention to the flaws during
their research, the companies operating data exfiltration services improved the
data collection practices.
As we highlight the unsolved
problem of collecting personal data reflected outside input elements, we
hope data-collecting companies like Hotjar will address the problem.

This paper is organised as follows. Section \ref{sec:background} provides the
necessary theoretical background on session replays, heat maps, options to
avoid being recorded, and other related work.
Section \ref{sec:methodology} describes the methodology used to analyse Hotjar
with achieved results in section \ref{sec:results}. Section~\ref{sec:discussion}
discusses our findings and proposes to extend the obligations of data protection
by design and default to processors. The paper is concluded in section \ref{sec:conclusion}.

\section{\uppercase{Background and Related Work}}
\label{sec:background}


A \emph{web session} is a series of requests and responses between a web server. Several techniques exist for maintaining a session state, but sessions are often bound to a cookie \cite{session}. Cookies consist of data stored in the browser that is sent with every request to the server and can be modified with each response.
Web pages track sessions by identifiers in cookies \cite{session}.

\emph{Session recordings} are usually used by web operators with the primary goal
of a better understanding of user behaviour~\cite{session_replay_comparison}. Consequently, web operators can improve the user
experience. Web operators can replay the users' interaction with the
website as a playback. Such playback can be analysed and annotated by web
operators. For example, the operators can identify sessions during which users
were confused or frustrated with a web page, analyse the behaviour, and improve
the page.

\subsection{Heat Maps}
A heat map visualises users' interaction with the
website. It uses colour intensity to demonstrate the clicks. The brightness of a
heat map reflects how popular a particular website section is.
Without displaying the data numerically, a heat map provides quick and simple-to-understand visuals \cite{kaur2015analysis}, facilitating the analysis and better understanding of how the user interacts with a~website.

\begin{figure}[ht]
    \centering
    \includegraphics[width=0.36\textwidth]{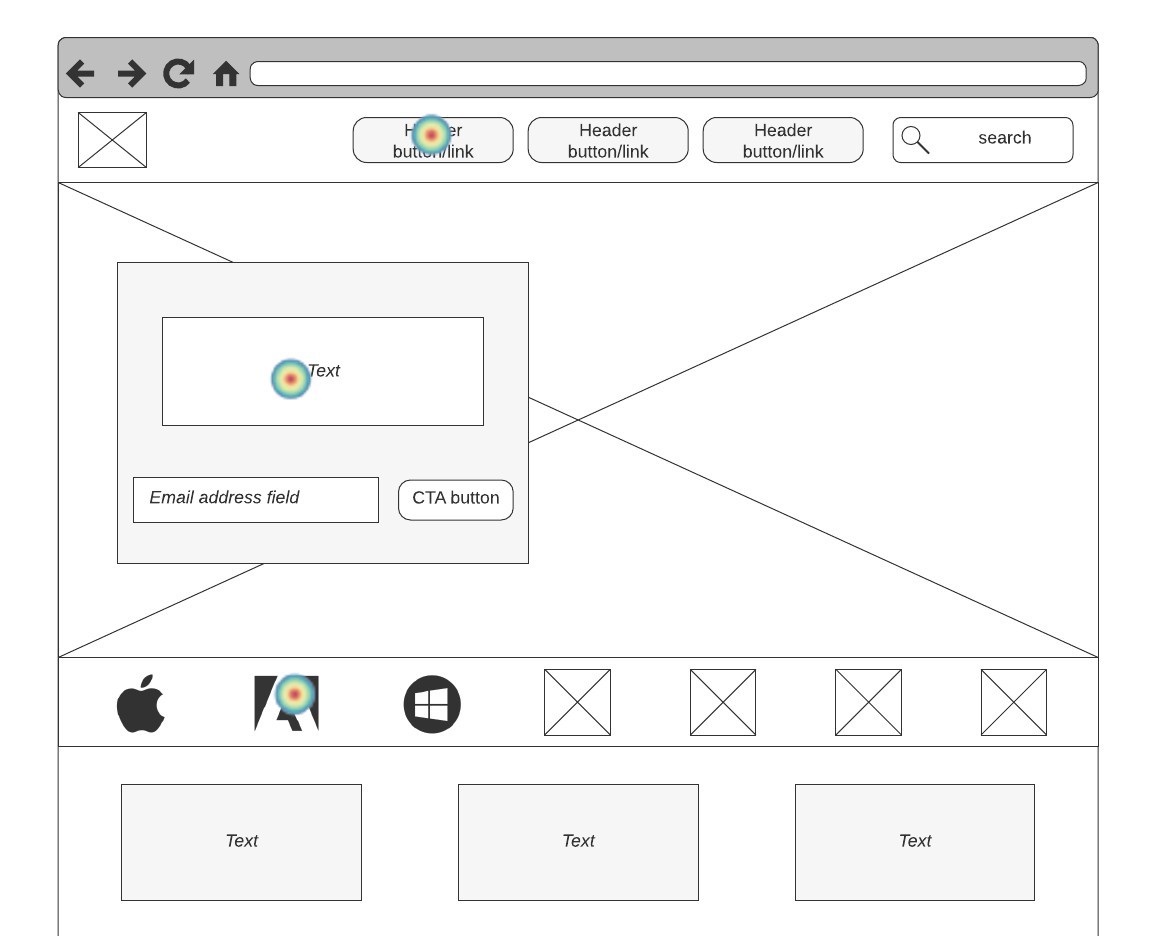}
    \caption{Click map.}
    \label{fig:click_map}
\end{figure}

\emph{Click heat maps} visualise areas where the user clicks the most, as shown in
Figure \ref{fig:click_map}. Click heat maps are derived from click coordinates,
target element, device resolution, and web page content. This information
reveals where users click and which element looks clickable \cite{session_replay_comparison}.

\begin{figure}[ht]
    \centering
    \includegraphics[scale=0.15]{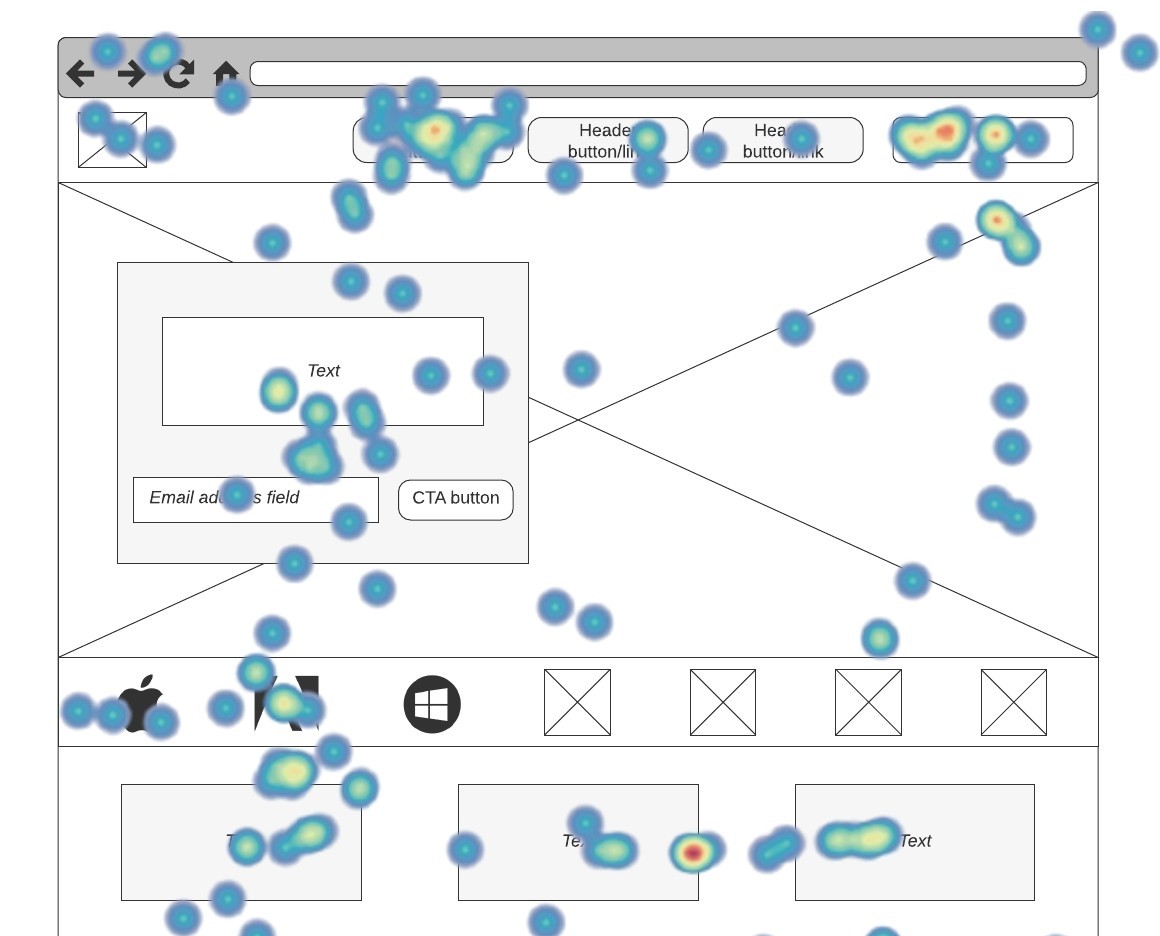}
    \caption{Mouse-tracking heat map.}
    \label{fig:mouse_movement}
\end{figure}

\emph{Mouse-tracking heat maps}, shown in Figure \ref{fig:mouse_movement}, are
similar to click maps, but
they visualise areas where users hover the most.

\begin{figure}[ht]
    \centering
    \includegraphics[scale=0.2]{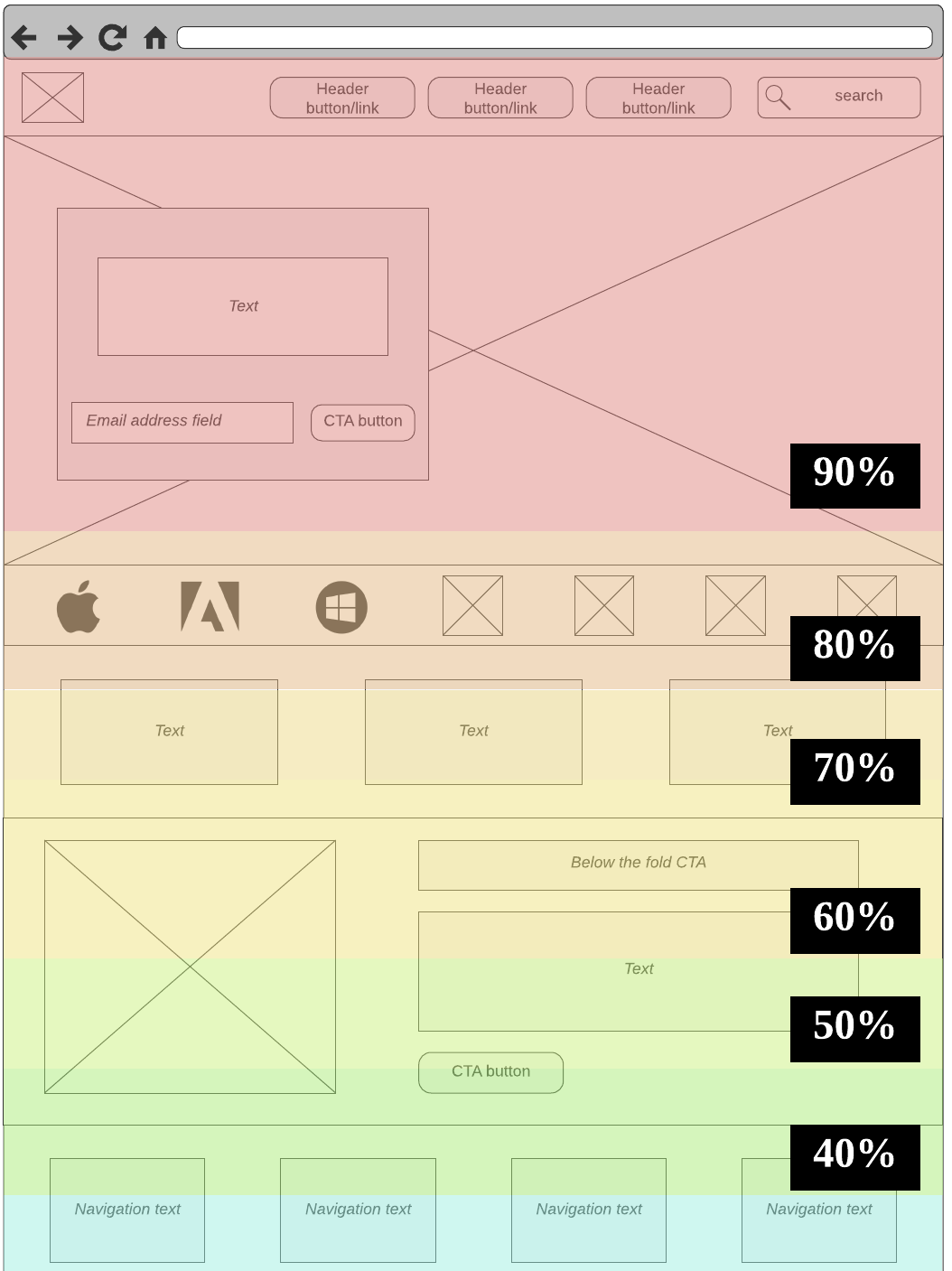}
    \caption{Scroll map.}
    \label{fig:scroll_map}
\end{figure}

\emph{Scroll heat maps} visually represent the user's scrolling behaviour. They provide
information on how far a user scrolls down on a website, see Figure~\ref{fig:scroll_map}.

\subsection{Session Replay Scripts}
\label{sec:session-replay-script}

A session replay script records the user's interaction with websites or
applications and sends the recordings to the server, where it is processed and converted to
a~format that can be replayed. Additionally, the server aggregates the
recordings and generates heat maps.

Session recording companies usually offer session replays and heat maps as a
service deployed by many websites. Both session replays and heat maps depend
on gathering data such as the content of the whole page, displayed text,
mouse movements and clicks, and key presses.
As the
user's device is identified by the generated tracking ID stored in a cookie,
web operators can monitor user behaviour over a long period \cite{session_replay_comparison}. 

\begin{figure*}[ht]
    \centering
    \includegraphics[width=0.7\textwidth]{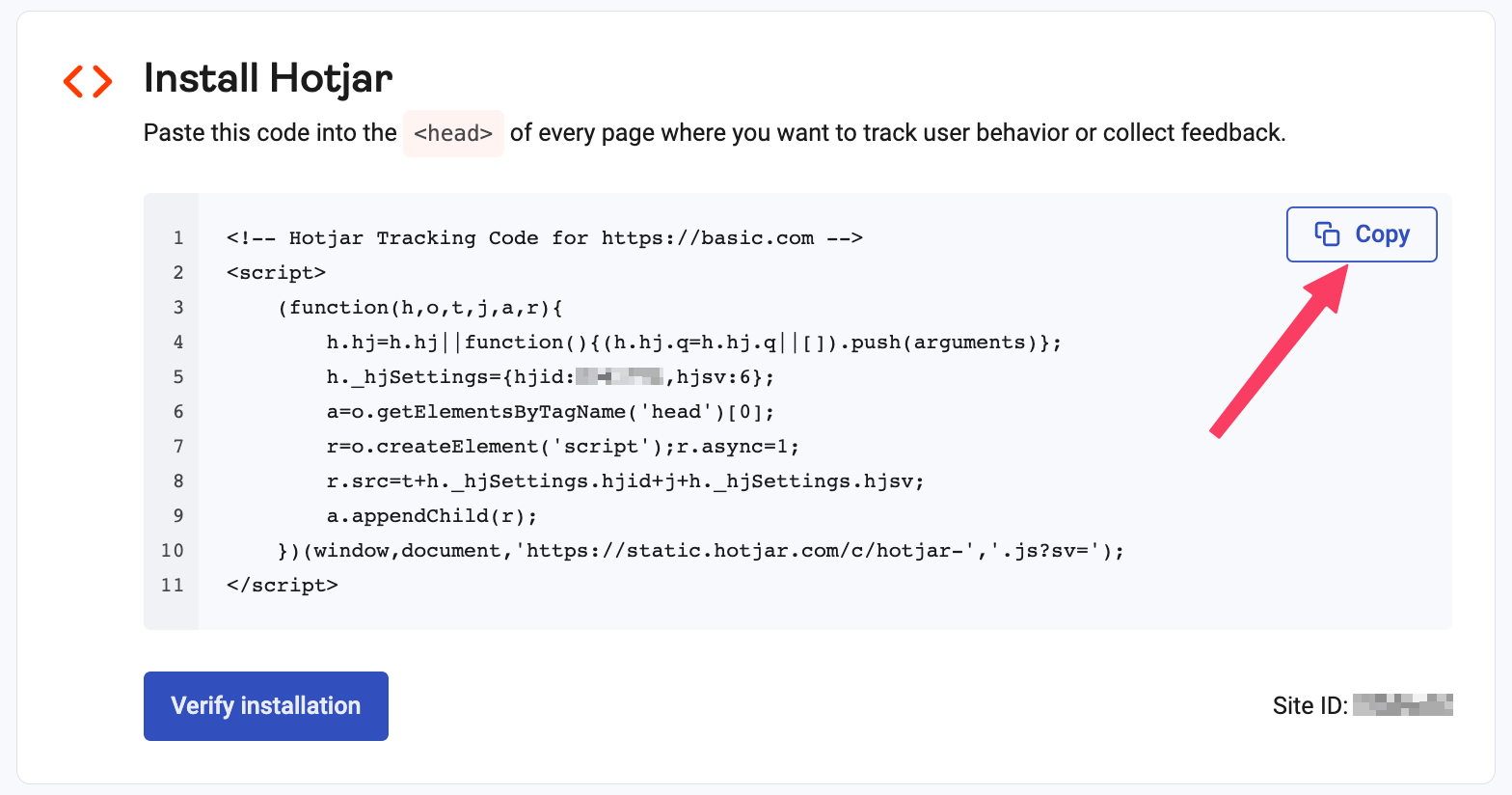}
    \caption{Hotjar installation script.}
    \label{fig:hotjar_installation}
\end{figure*}

Web operators can install session replay scripts easily. Session recording
companies market themselves as being easy to deploy \cite{session-replay}. For instance, Hotjar provides
the instructions depicted in Figure \ref{fig:hotjar_installation}. This code
inserts another script into the website. The page environment is initialised to
set up the data gathering.

However, users are not typically aware of the detailed monitoring of their
activities~\cite{session-replay,session_replay_nissenbaum}. Such data collection practice has been particularly problematic in
cases involving sensitive data unless the application developer has manually
redacted the website or application to ensure the user's
privacy~\cite{session-replay}. Gathered data can be analysed to derive heat
maps, recordings of mouse movements, or the success of filling forms.

\subsection{Privacy Impact Analysis}

Several studies have already analysed the impact of the data exfiltration from
web pages with recording scripts. Passwords, credit card numbers, and health data do leak to third
parties, often without any awareness by the website
owner~\cite{session-replay,leaky_forms_usenix}.
Pages like contact forms and registration and authentication forms leak data that third parties can
use to de-pseudonymise web site
users~\cite{formlock,registration_data_leaks,authentication_data_leaks}.
Companies perform such de-pseudonymisation, for example, to track users on
different devices~\cite{ftc_cross_device} as such identifiers are stable and
persistent~\cite{leaky_forms_usenix,authentication_data_leaks}.

Worse, session replay scripts and other trackers that leak private information
are omnipresent~\cite{session-replay}. At least some of the session replay
script vendors leave the responsibility to configure web sites to their
customers~\cite{session_replay_nissenbaum}. Nevertheless, previous research
identified that web sites leak private information. For example,
\cite{Krishnamurthy2011} reported that 56\,\% of sites leak private
information, and \cite{authentication_data_leaks} discovered that 42.3\,\% of sites
leak authentication data to third parties. Other research reports a lower share of
leaking sites~\cite{registration_data_leaks}.

Previous research shows that trackers are less frequent in the EU compared to
the US~\cite{leaky_forms_usenix}. Studies of privacy notices show that they are
not
clear~\cite{authentication_data_leaks,ftc_cross_device,registration_data_leaks}.

Although \cite{session_replay_nissenbaum} considered malicious web operators
from the ethical and legal point of view, to our best knowledge, we are the
first to study technical ways that mislead the collectors to gather data that
are not recorded by the unmodified session replay script.

\subsection{Ways to Avoid Tracking}
\label{subsec:dnt}

Tracker blockers employ lists of URLs or parts of URLs
that are considered harmful to user privacy or security. The advantage for the user is
that many tools focus on blocking (for example, uBlock Origin\footnote{\url{https://github.com/gorhill/uBlock#ublock-origin}}, EFF
Privacy Badger\footnote{\url{https://privacybadger.org/}},
Ghostery\footnote{\url{https://www.ghostery.com/}}) and blocklists are usually
compatible with several blockers. Browsers like Firefox~\cite{FirefoxTP} and Brave include
tracking prevention by default. Previous research shows that by blocking
trackers, users improve the performance of their browsers~\cite{FirefoxTP}.
The downside of the list-based blockers is that blockers can evade detection
by changing the script URL so that the rule
in the block list no longer matches the URL of the
tracker~\cite{block_me_if_you_can}. Worse, block lists contain only
previously detected trackers and, by definition, cannot contain tracking domains
in advance. Consequently, researchers find tracking domains that are not a part
of any popular block list~\cite{leaky_forms_usenix}.

Do Not Track (DNT) is a web browser setting that allows users to signal opt-out
from tracking. DNT aims to
enable users to communicate their tracking preferences to each server. A user's
web browser sends an HTTP header called DNT. Once a server receives the
\texttt{DNT: 1} header, it should stop tracking the user.
\texttt{DNT: 0} means that the user prefers to allow tracking. DNT also
allows access to the preference via JavaScript API
\texttt{navigator.doNotTrack}.
DNT was supposed to become a W3C standard \cite{w3cdnt}.
Nevertheless, websites do not
generally respect the signal~\cite{privacy_signals}, and the standard is retired
due to a lack of enforcement and minor effects. Even so, browsers still allow
users to activate DNT.

DNT signal received a little success~\cite{privacy_signals}.
\cite{session-replay} did not study if Hotjar and other session collecting
companies respect DNT. Moreover, \cite{session_replay_nissenbaum} claim that
users typically do not have the option of opting out. Nevertheless, Hotjar mentions respect for DNT in its
privacy statements. We validated the code and confirmed that Hotjar indeed
respects DNT. Even so, we warn that malicious website operators or attackers
can circumvent the DNT detection and force Hotjar to collect data from users
signalling their no-tracking preferences.

\section{\uppercase{Methodology}} \label{sec:methodology}

We reproduce the tests mentioned in the original
study~\cite{session-replay} and determine if
something has changed in Hotjar to ensure users' privacy.

Firstly, we studied Hotjar policies and documentation and opened a testing account.
We validated that Hotjar respects the Do Not Track
signal (Section~\ref{subsec:dnt_methodology}). To do so, we read the Hotjar
script and located the code responsible for Do Not Track handling.
Next, we checked if the interaction with
the testing account was encrypted (Section~\ref{subsec:tls_results}).

Secondly, we analysed the nature
of collected user data with a focus on dynamically generated data by both the
user and the page as a result of the users' interaction with the page
(Section~\ref{subsec:tranferred_data}).

In parallel, we investigated the possibility
of the web server operators modifying the Hotjar script to increase the amount of
collected data (Sections \ref{subsec:dnt_methodology}--\ref{subsec:tranferred_data}).

We created three web applications:

\begin{enumerate}

  \item A log-in form with a password input and text input for email username.
    As many log-in forms allow a user to show the verbatim password, we included a
    button that changes the input type of the password input to text. The goal
    was to (1) compare data collected from password inputs to other input types
    and (2) validate that passwords are treated consistently if the user decides
    to peak into the field.

  \item A shipping form where a user needs to fill in name, address, email, phone,
    credit card details, and other information that is typically needed for
    shipping. The goal was to study other input types. Hence, the page reflects
    the content of an input element in other elements.

  \item A page that changes the content according to the user input. The page
    contains a text input. The page offers country
    names matching the input text as the user types text in.

\end{enumerate}

We created the log-in and shipping form in three varieties:

\begin{enumerate}

  \item The first scenario follows Hotjar's
    documentation. ID attributes of both input fields were set to the
    appropriate values: \texttt{pass} or \texttt{password} for the
    password input field and \texttt{email} for the email input field.
    According to documentation, Hotjar
    should not record such input fields. Instead, it should replace their content with
    asterisks of arbitrary length. The shipping form also employs specific ID attributes
    to prevent the potential leakage of sensitive personal information.

  \item Even though Hotjar does not record keystrokes by default, the web operator can
allow collecting the content of any input element by adding attribute
\texttt{data-hj-allow}. Consequently, Hotjar records the content that the user
    inserts into such fields verbatim and does not replace it with asterisks.
    The form did not use any well-known ID for the inputs. Instead, we
    chose IDs randomly.

  \item The two scenarios above use the original Hotjar session recording script.
    The third scenario is the same as the first scenario, but we modified
    the Hotjar script. For example, we modified the function that masks the
    username and password. The original function replaces the real text with a
    random number of asterisks. The modified version does not
    transform the original string in any way.

\end{enumerate}

Finally, we tested the default configuration of multiple browsers and validated
if the built-in protections prevent pages from recording the sessions, see
section~\ref{subsec:browsers_protection}.

\section{\uppercase{Results}} \label{sec:results}

This section provides the results of the experiments designed in
Section~\ref{sec:methodology}.

\subsection{Respecting DNT} \label{subsec:dnt_methodology}

Unlike websites that refuse to respect DNT signals~\cite{privacy_signals},
Hotjar script reads \texttt{navigator.doNotTrack} property, see
Figure~\ref{fig:hotjar_dnt_script}. When the
property carries the value 1, the recording of a particular session is omitted.
Nevertheless, browsers with DNT activated still need to download and execute the Hotjar
script as the check is not part of the code inserted directly to the web page
using Hotjar services (the code in Figure~\ref{fig:hotjar_installation}).

\begin{figure}[ht]
    \centering
    \begin{verbatim}
"1"!==navigator.doNotTrack&&
"1"!==window.doNotTrack&&
"1"!==navigator.msDoNotTrack
    \end{verbatim}
    \caption{The script at
    \url{https://script.hotjar.com/modules.1e98293c16a88afdf1b7.js} gathers data
    only if it does not detect DNT.}
    \label{fig:hotjar_dnt_script}
\end{figure}

The retired DNT standard allows specifying extensions to the
\texttt{navigator.doNotTrack} property~\cite[Section 5.2.1]{w3cdnt}. The clause
in Figure~\ref{fig:hotjar_dnt_script} would not detect such extensions, and the
preference not to be tracked would not be honoured in the presence of extensions
(such as \texttt{"navigator.doNotTrack == 1xyz"}). Nevertheless, the retired standard warns against
using extensions, so hopefully, DNT implementers follow the suggestion not to set
extensions.

A malicious website can change the value
of the \texttt{doNotTrack} property to record all sessions, for example, by
executing the code in Figure~\ref{fig:dnt_set_tracking} before including the
Hotjar script. Although browsers still send the \texttt{DNT: 1} HTTP header,
Hotjar captures the recording. Thus, a malicious site can confuse Hotjar scripts
to record sessions of users opting out of tracking.

\begin{figure}[ht]
    \centering
    \begin{verbatim}
Object.defineProperty(navigator,
                     "doNotTrack", {
        get: function () { return "0"; },
        set: function (a) {},
        configurable: false
      });
    \end{verbatim}
    \caption{A malicious website can reconfigure browsers to allow tracking by
    Hotjar.}
    \label{fig:dnt_set_tracking}
\end{figure}

\subsection{TLS Support And Dashboard Data Encryption}
\label{subsec:tls_results}

Since user data ends in the session recording, recording services must
prioritise security. Otherwise, the protection of user data may fail. Hotjar
used to deliver playbacks within an HTTP
page\footnote{\url{https://freedom-to-tinker.com/2017/11/15/no-boundaries-exfiltration-of-personal-data-by-session-replay-scripts/},
section 4: \emph{"The publisher dashboards for Yandex, Hotjar, and Smartlook all
deliver playbacks within an HTTP page, even for recordings which take place on
HTTPS pages."}}, even for recordings on HTTPS pages. This allowed a
man-in-the-middle to insert a~script into the page to extract all the data from
the record. Today, Hotjar uses HTTPS.

Moreover, the Hotjar script refused to work when deployed on web site without TLS.
Hence, a man-in-the-middle adversary cannot misuse Hotjar to record sessions by
adding the Hotjar script to HTTP web pages.

\subsection{Transferred Data} \label{subsec:tranferred_data}


Hotjar records the position and timestamp of each click or mouse hover and
entered data in input fields, including the time and the timestamp of the input,
selector, and input field type. Hotjar also records location, operating system and
resolution. Nevertheless, recording of the user's location can be disabled in
Hotjar settings by the web operator.

\subsubsection{Usernames and Passwords}
The original study \cite{session-replay} claims that passwords are excluded from
the recordings to prevent password leaks. To validate this statement, we created a sign-in form with two input fields of type email and password and tested different scenarios mentioned in section \ref{sec:methodology}. Here, we provide the test results for each scenario.

Hotjar collecting script masks
both inputs (username and password) of the log-in form, following Hotjar
guidelines. The Hotjar session playback displayed
asterisks in both input fields. The number of asterisks is different from the
actual password length. Hence, the replay does not leak the real length of the
password.

The other log-in form contains the password input element. Both input elements for username
and password are decorated with \texttt{data-hj-allow} attribute. Even so,
Hotjar does not record the content of both input fields. However, once the user
interacted with the \textit{show password} feature, our test website changed the
type of input fields, causing the user's password to become visible. In this
case, Hotjar failed to mask the password field, resulting in capturing the
password in the clear.

As expected, the modified script that did not replace the input text let the
entered data leak to Hotjar. The usernames, as well as passwords after using
the \textit{show password} feature, were then available to the website operator
through the Hotjar dashboard.

\subsubsection{Shipping details}

Firstly, all entered data to the form created according to Hotjar guidelines to
prevent leaks of personal information were replaced with asterisks. Credit
card information was replaced with asterisks no matter the data type of the
input. Moreover, the phone number was replaced with ``1111''.

Since most input fields in the shipping form are of the text type, web operators
can allow data collection using \texttt{data-hj-allow} attribute. Such change
allows the web operator to collect first and last names, phone numbers, company
names, and credit card information.

\cite{session-replay} warned that the \texttt{autocomplete} attribute of HTML
input elements could leak user passwords outside the log-in process without the
awareness of the user. The Hotjar recording script does not
mask the content of input elements with both \texttt{data-hj-allow} and
\texttt{autocomplete} attributes.

The modified session replay script allows adversaries to record all entered data, including credit card information.

\subsubsection{Rendered Website Content}

Hotjar collects rendered page content. Unlike user input recording, the
collecting script does not suppress the rendered content unless redacted
manually by \texttt{data-hj-suppress} attribute. The default configuration
leaks all displayed content in our tests. In the testing form
with one input field, the script did not collect the content of the input field
as it lacked the \texttt{data-hj-allow} attribute. However, the search results
leaked into the recordings. This allows one to guess the content of the input search field.

\subsection{Protections Offered by Browsers}
\label{subsec:browsers_protection}

Section \ref{subsec:dnt} describes how to change the read-only
\texttt{navigator.doNotTrack} property, which lets Hotjar inject the
recording scripts. The session was recorded using web browsers such
as Google Chrome, Microsoft Edge, Firefox or Opera. However, not every session
was recorded. Firefox contains tracking protection that blocks
content loaded from domains that track users. Opera also offers tracking protection
called Tracker Blocker, which blocks the Hotjar session replay script.

\section{\uppercase{Discussion}} \label{sec:discussion}

Article 29 of the EU Directive 95/46/EC established The Article 29 Working Party
(WP29). GDPR transformed WP29 into the European Data Protection Board (EDPB) with
increased powers. Both WP29 and EDPB publish guidelines and opinions with the
aim of consistent application of data protection law. WP29 published an opinion
on the ePrivacy Directive~(Directive 2009/136/EC) consent exception. By applying
the opinion on session replay scripts, it is clear that users need to consent
to session recording whenever the ePrivacy Directive applies. Even so, the code
displayed in Figure~\ref{fig:hotjar_installation} does not seek any consent.
Consequently, the code may violate European data protection law.

The number of users that signal the DNT flag is not known. However, Mozilla
reported that approximately 11\,\% of global Firefox users and approximately
17\,\% of US users signal DNT \cite{mozilla_dnt_share}. Today, the
share of users is likely different. Hence, it is difficult to estimate the impact on
the privacy of web users.

The shift to TLS is likely affected by the global adoption of
TLS~\cite{lets_encrypt}. Recordings no longer transit
Hotjar servers in the clear. Moreover, Hotjar disabled its scripts at HTTP-only sites,
further stimulating global HTTPS adoption. 

Distinguishing user-specific non-input elements on the page will likely
prove difficult. As explained in Section~\ref{sec:session-replay-script} and
illustrated in Figure~\ref{fig:hotjar_installation}, the ease of installation is
one of the goals of Hotjar. Since non-input elements typically do not contain
user-specific data, collecting all data by default makes sense.
The controller can add \texttt{data-hj-suppress} attribute to disallow data
collection from elements containing user-specific content.

However, as \cite{session-replay} highlight in the economic analysis of the
failures, web operators lack the budget to hire experts to annotate elements
that should not be collected. \cite{gdpr_costs} studied the costs
of complying with GDPR, the height and the likelihood
of fines: small and middle-sized companies should only pay
a few cents or euros for GDPR legal compliance. Hence, web operators are not
motivated to launch audits on their data collection.

Hotjar can treat all HTML elements to have \texttt{data-hj-suppress} by default,
which would replace all generated and reflected content with asterisks at the cost of
removing all static content. That would motivate web site operators
to clearly mark texts that are safe to be recorded. We see two risks with this
approach: (1) it makes the deployment of session replay scripts
more time-consuming, and (2) the web operators might take the easy way and mark
all elements with the \texttt{data-hj-allow} attribute. Even though such steps
might be illegal, the risk of a fine is so small~\cite{gdpr_costs} that such
behaviour might appear in practice.

\cite{session-replay} considered potential violations of laws like GDPR in the
EU and HIPAA (healthcare) and FERPA (education) in the United States. To their
best knowledge, courts have not decided claims concerning processorship and
joint-controllership of session recording companies. In contrast, we believe
that the rulings of the Court of Justice of the European Union (CJEU) on the concept of
data controllers and processors~\cite{cjeuC210_16,cjeuC40_17} are applicable to
the case of session recording companies. EDPB published guidelines following the
judgments~\cite{edpb_controller}.
As long as the session recording companies follow the instructions of the
website operators, they can be assumed processors.

As it is the operator that needs to explicitly enable data collection from input
elements with the \texttt{data-hj-allow} attribute, we believe it is the web
operator who controls the data collection, and Hotjar is a processor.

Article 25 of GDPR mandates that personal data controllers apply data protection by
design and default. However, it does not prescribe the same obligations to
processors. As numerous web page operators (controllers) deploy
session recording scripts, but the number of session recording vendors is
small~\cite{session-replay}, legislators should consider extending Art. 25
obligations to processors. We believe that such requirements would remove the
option of session replay vendors shifting the responsibility to web operators
when it is not a sincere expectation that they would hold to their
responsibilities~\cite{session_replay_nissenbaum,gdpr_costs}.

Section~\ref{sec:results} explains several ways how malicious web site operators
can modify the data-collecting script.  We expect that the web site operators
would be liable for such misconduct. However, that might prevent session replay vendors
from applying technical measures to prevent unauthorised data collection.

\cite{session_replay_nissenbaum} highlights that the need to deploy session
replay scripts is decreased when following best practices to design UI, and their
deployment should be as narrow as possible and with clear objectives.
CJEU clarified the meaning of strict necessity~\cite[recital
46]{cjeuC708_18} that would likely need to be fulfilled to make data collection
legal in the European Union without the consent of data subjects.

\section{\uppercase{Conclusion}} \label{sec:conclusion}

\noindent Session recordings allow web operators to detect problems and
optimise websites with a limited budget. However, extensive data collection
practices can violate laws like the ePrivacy Directive, HIPAA, and FERPA.
This position paper focused on Hotjar, a session recording company
originally investigated by \cite{session-replay} several years ago.

We show that
the privacy of users on websites with Hotjar scripts improved. For example,
the content of input elements is no longer collected by default as it typically
contains user-specific data.
Nevertheless, we point out several possibilities of
adversaries trying to circumvent the protections of Hotjar. User-specific data
stored outside input elements are still collected by Hotjar by default; even
though we suggest the application of the \texttt{data-hj-suppress} attribute to all
DOM elements by default, we
argue that data collection of reflected content will likely continue. We also argue that
European data protection law likely makes web site operators liable for the
issues. However, the price of a data collection audit is several magnitudes
higher than the expected costs of non-compliance for almost all companies.

Future research should (1)~validate and compare other session recording
companies, (2)~study the presence of recording without consent, (3)~study
the liability of session recording companies for recordings without consent, and
(4) propose incentives to audit data collection on websites.

\section*{\uppercase{Acknowledgements}}

This work was supported by the Brno University of
Technology grant Smart information technology for a resilient society
(FIT-S-23-8209).

\bibliographystyle{apalike}
{\small
\bibliography{biblio}}


\end{document}